\input{aipcheck}

\documentclass[
    ,final            % use final for the camera ready runs
  ]
  {aipproc}

\layoutstyle{6x9}
\begin{document}

\title{An improved redshift indicator for Gamma-Ray Bursts, based on the prompt emission}



\author{A. P\'elangeon}{
  address={Laboratoire d'Astrophysique, Observatoire Midi-Pyr\'en\'ees, 31400 Toulouse, France}
}

\author{J-L. Atteia}{
  address={Laboratoire d'Astrophysique, Observatoire Midi-Pyr\'en\'ees, 31400 Toulouse, France}
}

\author{D.~Q. Lamb}{
  address={Department of Astronomy \& Astrophysics, University of Chicago, Chicago, IL 60637, USA}
}

\author{G.~R. Ricker}{
  address={Center for Space Research, Massachussetts Institute of Technology, Cambridge, MA 02139, USA}
}

\author{the HETE-2 Science Team}{
  address={An international collaboration of institutions in USA, France, Japan, Italy, Brazil and India}
}

\begin{abstract}
We propose an improved version of the redshift indicator developed
by Atteia \cite{Atteia03}, which gets rid of the dependence on the
burst duration and provides better estimates for high-redshift GRBs.
We present first this redshift indicator, then its calibration with HETE-GRBs with known redshifts. We also provide an
estimation of the redshift for 59 bursts, and we finally discuss the
redshift distribution of HETE-bursts and the possible other
applications of this redshift indicator.
\end{abstract}

\maketitle

%%%%%%%%%%%%%%%%%%%%%%%%%%%%%%%%%%%%%%%%%%%%
%% MAINMATTER
%%%%%%%%%%%%%%%%%%%%%%%%%%%%%%%%%%%%%%%%%%%%

\section{Description}
In 2003, Atteia proposed
$X_{0}=N_{\gamma}/(E_{peak}\times\sqrt{T_{90}})$ as a possible
redshift estimator \cite{Atteia03}, based on the $E_{peak}-E_{iso}$
correlation \cite{Lloyd,Amati02} linking $E_{peak}$, the intrinsic
peak energy of the $\nu f_{\nu}$ spectrum, and $E_{iso}$, the
isotropic energy radiated by the source in its rest frame. In
addition, Yonetoku et al. have shown that the $E_{peak}-L_{iso}$
correlation was less dispersed than the $E_{peak}-E_{iso}$ correlation
(2004) \cite{Yone}.\\
The definition of our new redshift indicator is partly based on
these two relations and is written as : \emph{$X = n_{15}/e_{p}$},
where \emph{$e_{p}$} is the observed peak energy and \emph{$n_{15}$}
the observed bolometric luminosity in units of photons and in the 15
sec. long interval containing the highest fluence. Thus, all the
burst spectra are now done on the same duration in the observer frame.\\
To compute this estimator, the burst spectra are fit with a Band
model \cite{Band} which gives us the spectral parameters : $\alpha$,
$\beta$, $E_{0}$ and the fluence in the energy range $[E_{1}-E_{2}]$
of the detector. Then, as described in the paper of Atteia
\cite{Atteia03}, the theoretical evolution of X with the redshift is
computed for a ``standard'' GRB ($\alpha = -1$, $\beta = -2.3$,
$E_{0} = 250~keV$), considering a ``standard'' cosmology
($\Omega_{m} = 0.3$, $H_{0} = 65~km.s^{-1}.Mpc^{-1}$, flat
universe), and finally the estimation of the redshift is deduced by
comparison between the value of X obtained for the GRB based on its
spectral parameters, and the theoretical evolution of X. In the
following we call this estimation \emph{new pseudo-redshift} (hereafter \emph{npz}).\\
In addition, errors on pseudo-redshifts are computed : considering
first the errors on the spectral parameters obtained with the fit,
1000 values of X are simulated, then 1000 \emph{npz} associated are
also calculated, and errors on \emph{npz} are so derived (the errors
presented hereafter are at 90\% confidence level).

\section{Calibration}
%\section{Testing the new redshift indicator}
%\subsection{Comparison with the previous estimator}
\begin{figure}[!htb]
\begin{minipage}{16cm}
\begin{center}
\includegraphics[height=.25\textheight]{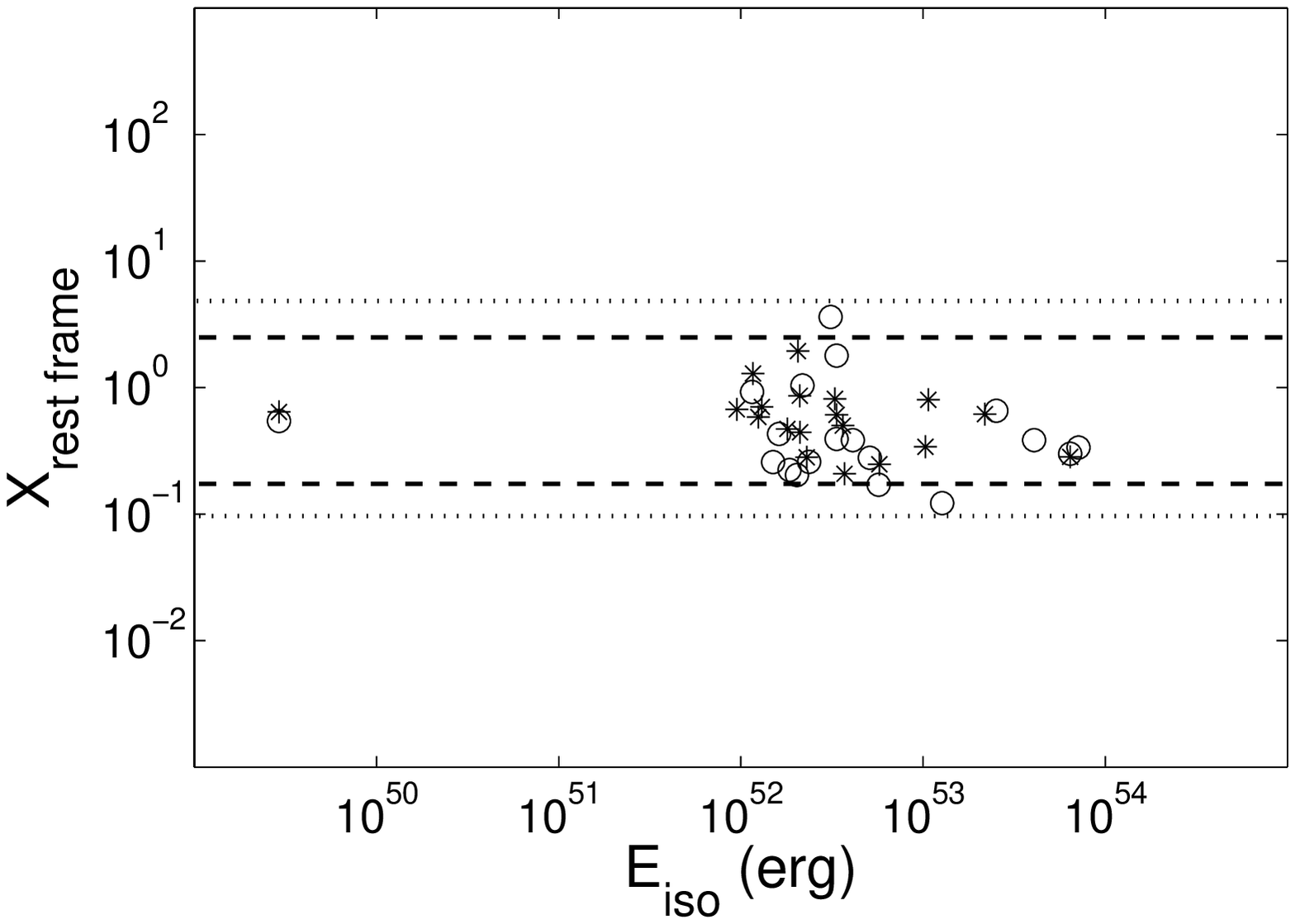}
\includegraphics[height=.25\textheight]{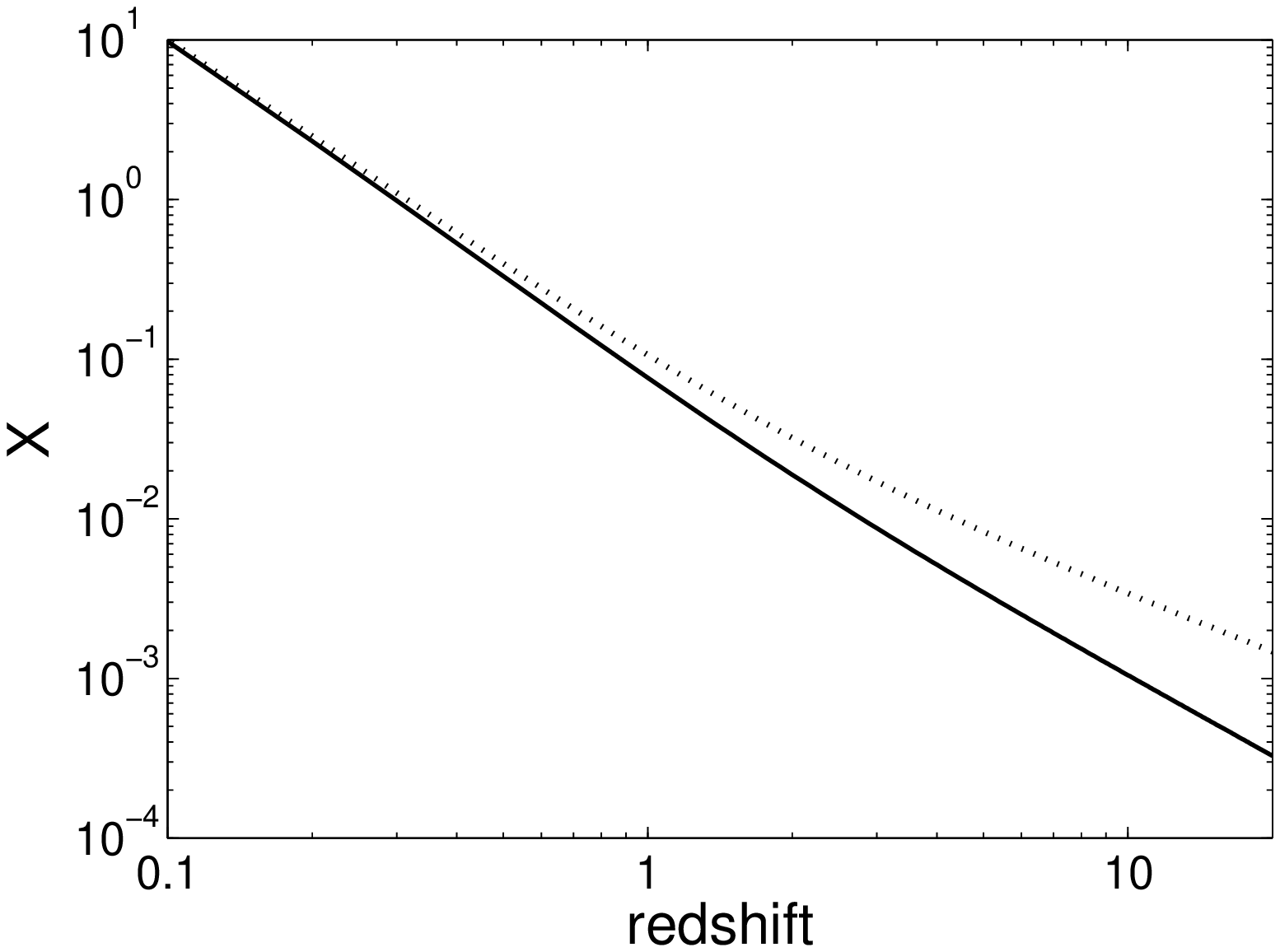}
\caption{Left panel : intrinsic dispersion of the two redshift indicators : 
$X_{0}=n_{\gamma}/(e_{p}\times\sqrt{t_{90}})$ \emph{(circles)} and $X = n_{15}/e_{p}$ \emph{(stars)} 
for 19 GRBs with spectroscopic redshift. Right panel : theoretical evolution of the two estimators 
\emph{(dotted line for the previous indicator,solid line for the new one)} between $z=0$ and $z=20$.} 
\label{intr} 
\end{center} 
\end{minipage}
\end{figure}
A good redshift indicator must essentially satisfy two criteria :
its independence on the bursts intrinsic characteristics, and its
high dependence on the redshift.\\Figure \ref{intr} \textit{(left
panel)} shows that whereas $E_{iso}$ is extended on about 5 decades,
the intrinsic dispersion of the new quantity X \textit{(stars)} is
only 1 decade against nearly 1.5 decade for the previous redshift
indicator \textit{(circles)}. In addition, the plot on the right
panel shows the higher dependence of the new redshift indicator
\textit{(solid curve)} with the redshift, where the difference
between the two estimators becomes significant for $z>1$.\\
%\subsection{Calibration}
\begin{figure}
\begin{minipage}{16cm}
\begin{center}
%%\includegraphics[width=5.5cm,height=7cm,trim=100 310 100 300,clip]{tableau1}
%%\includegraphics[width=5.5cm,height=5cm,angle=0,trim=10 5 20 15,clip]{npzsurz}
%% trim=gauche-bas-droite-haut
\includegraphics[height=.25\textheight,trim=100 310 189 320,clip]{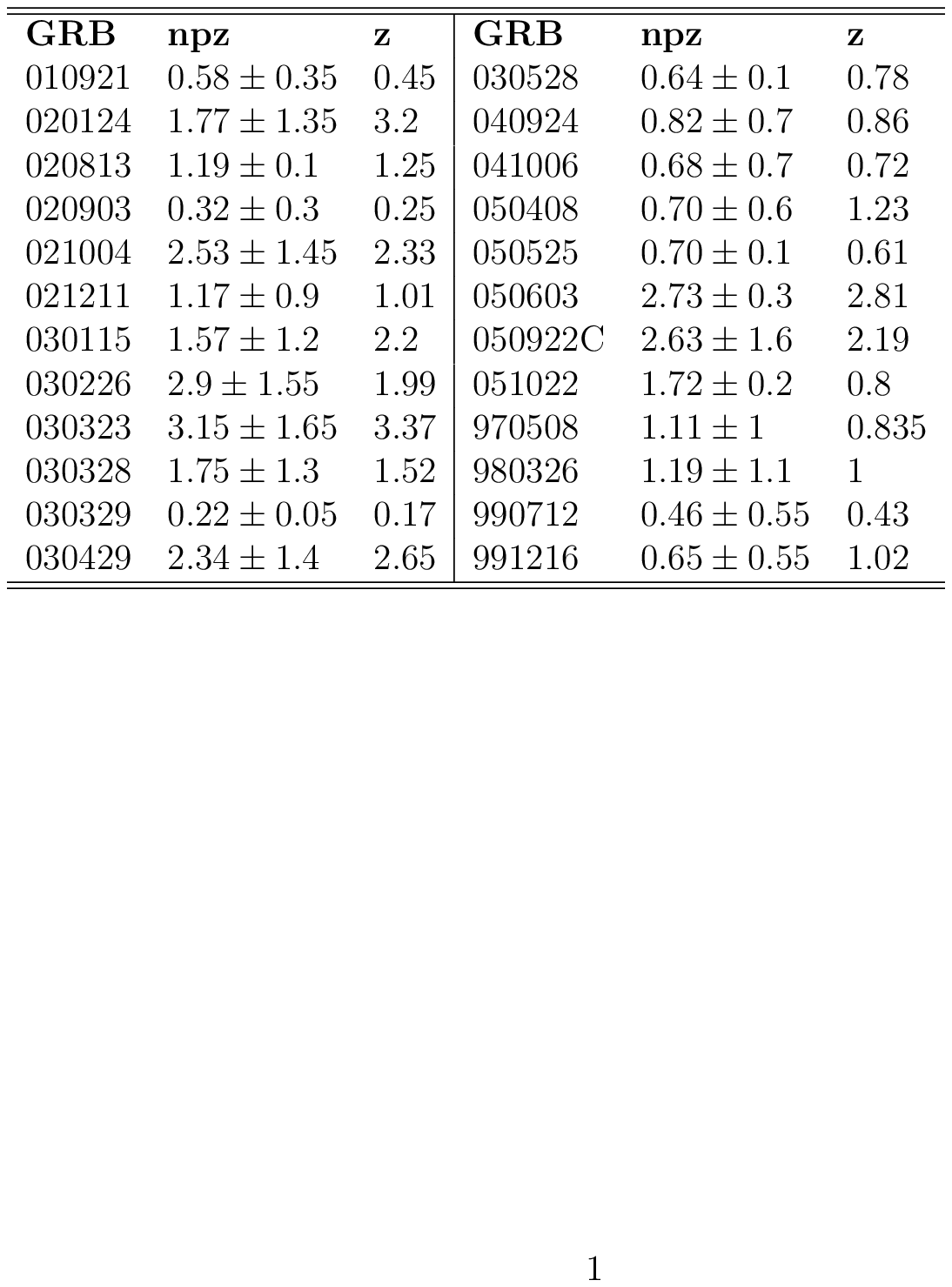}
\includegraphics[height=.25\textheight]{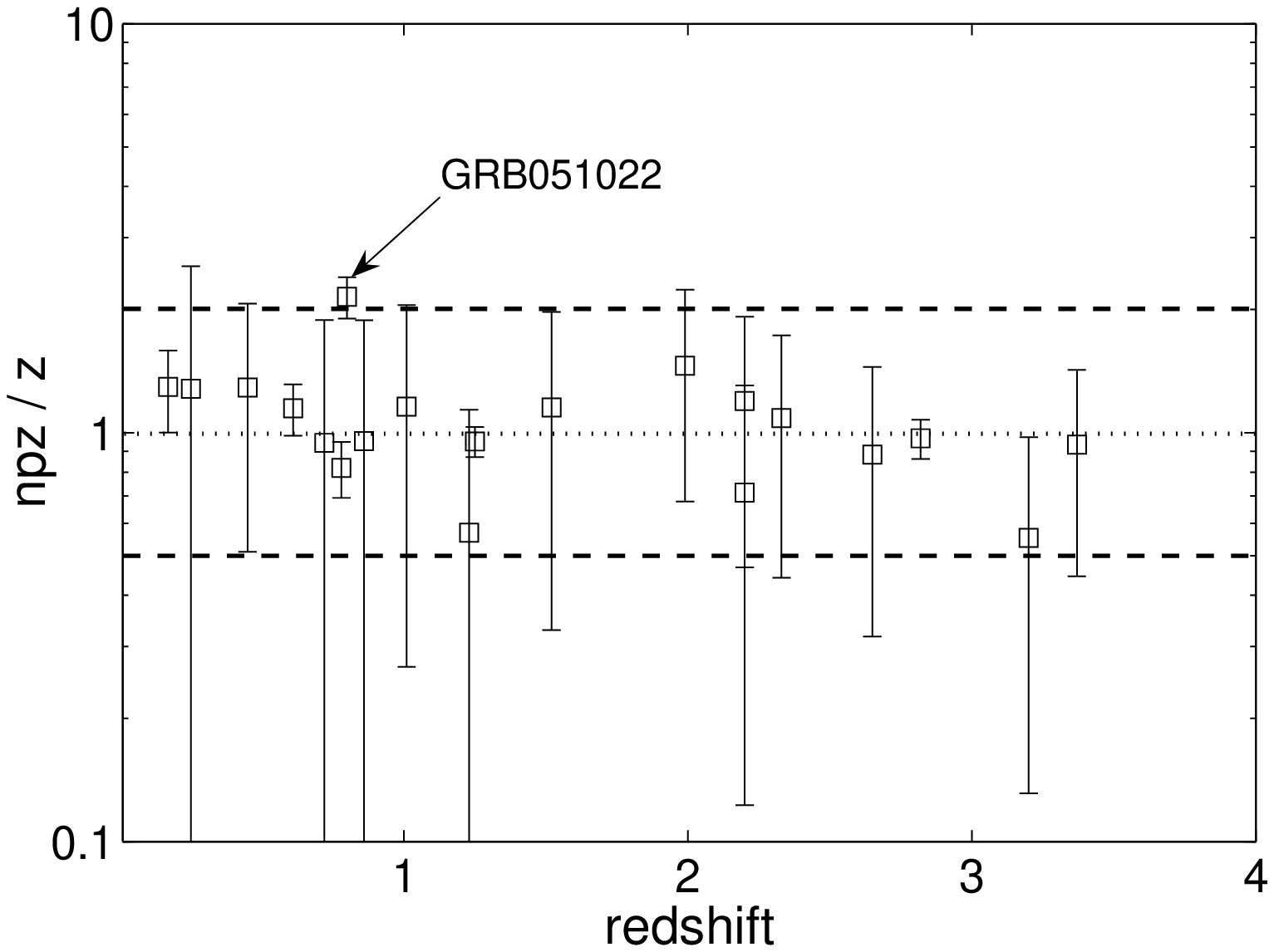}
\caption{Left panel: redshift estimates for 24 bursts with known spectroscopic redshifts. Right panel: ratio npz/z for 20 bursts with known redshifts.}
\label{tab1}
\end{center}
\end{minipage}
\end{figure}
The redshift indicator is currently calibrated with 17 GRBs detected
by \emph{HETE-2} which have a spectroscopic redshift, and 2
additional GRBs (\emph{050525} and \emph{050603}) detected by
\emph{Konus-Wind} \cite{GCN1}, \cite{GCN3}. The table on the
figure \ref{tab1} presents the results of the \emph{npz} obtained
with errors. We can notice \textit{(right panel)} that the redshift
estimate is always better than a factor 2 (\textit{dashed lines}),
except for \emph{GRB051022} which has a factor 2.15 and a small
error \textit{(see the arrow)}. We have yet to understand why this
burst seems to be an outlier. Without this last GRB, the standard
deviation is : $\sigma = 0.11~dex$. We have also computed the
pseudo-redshift for 4 other bursts referenced in the literature
\cite{Amati02,Ghirl} : \emph{GRB970508, 980326, 990712} and
\emph{991216}, which have a duration similar to the one used
($\sim15 sec.$) in the derivation of the npz.

\section{Study with the npz}
\subsection{A sample of bursts without spectroscopic redshift}
\begin{figure}[!htb]
\begin{minipage}{16cm}
\begin{center}
%%\includegraphics[width=7cm,height=8.5cm,trim=105 240 70 80, clip]{tableau2}
%%\includegraphics[width=5.5cm,height=5cm,angle=0,trim=10 7 55 25,clip]{cdfposter}
%% trim=gauche-bas-droite-haut
\includegraphics[height=.25\textheight,trim=100 270 240 280,clip]{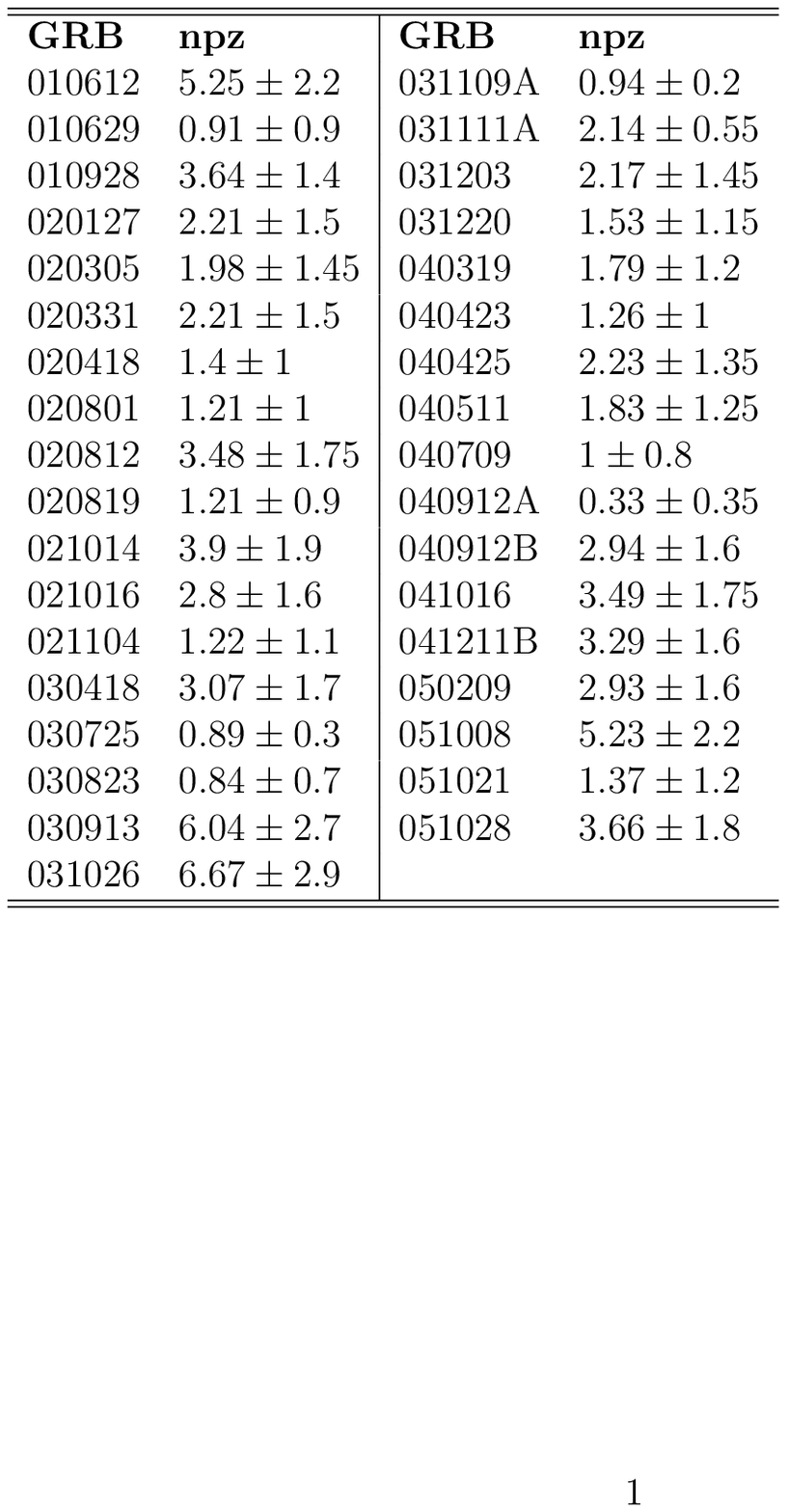}
\includegraphics[height=.25\textheight]{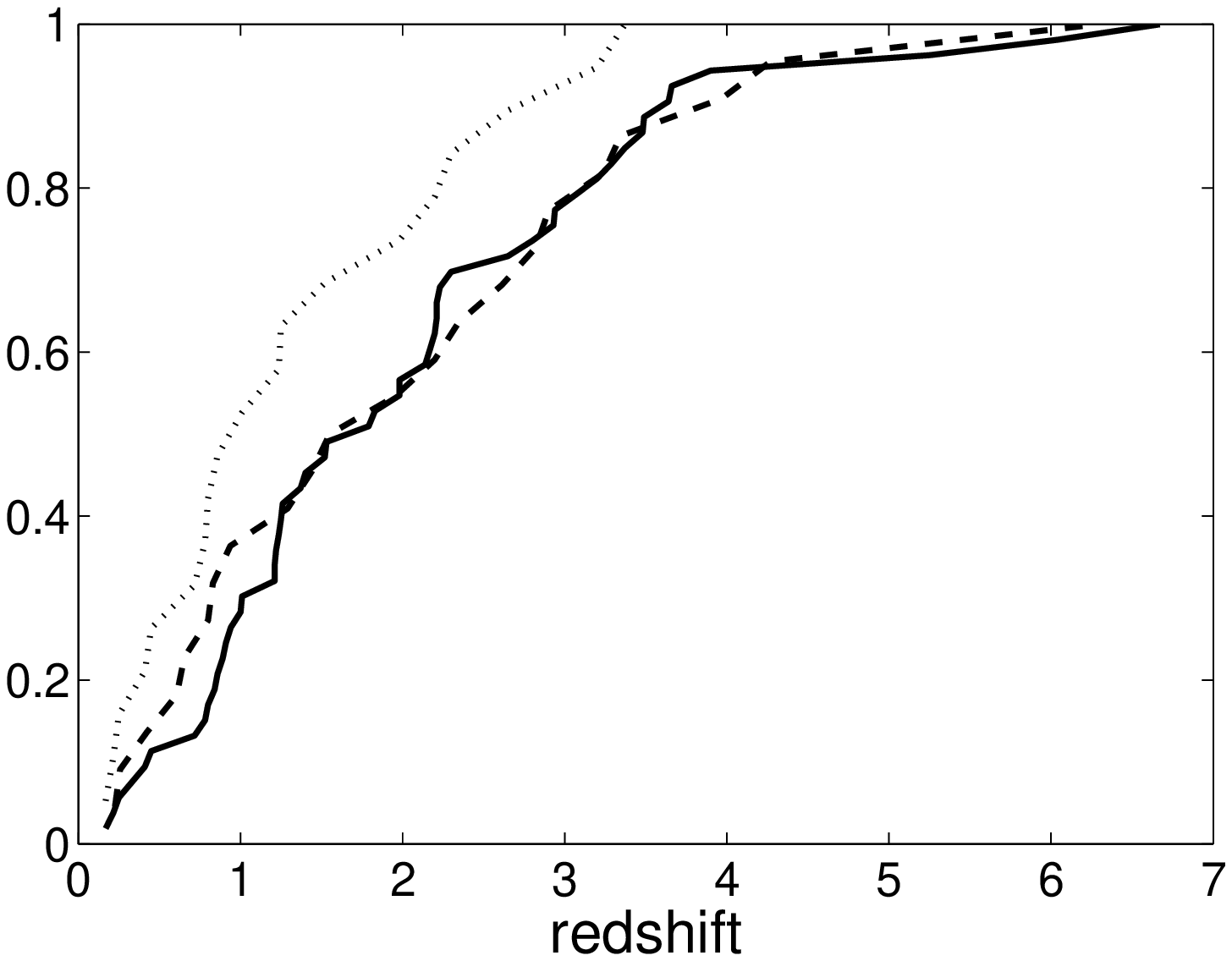}
\caption{Left panel : redshift estimates for 35 bursts without redshift (sample of 34 \emph{HETE}-bursts
and \emph{GRB051008} \cite{GCN4}). Right panel : cumulative distribution functions for
\emph{HETE}-bursts with redshift \textit{(dotted line)}, \emph{SWIFT}-bursts with redshift \textit{(dashed
line)} and a sample of 53 \emph{HETE}-GRBs with redshift or pseudo-redshift \textit{(solid line)}.}
\label{tab2}
\end{center}
\end{minipage}
\end{figure}

Taking into account all the long GRBs detected by the \emph{FREGATE}
instrument (6-400 keV) on-board of the satellite \emph{HETE-2} which
don't have spectroscopic redshift, we have computed a
redshift-estimate for a sample of 34 bursts which had enough
statistics to be correctly fit with \emph{FREGATE} data. The results
are given in the table presented in figure \ref{tab2}.\\
The GRB redshift distribution is probably biased because of the
small fraction of bursts which have a measure of their redshift. We
tried to determine if this fact is confirmed for the \emph{HETE-2}
bursts, and what could be this distribution if we had a higher
fraction of GRBs with known redshifts. Thus, we considered the
redshift distribution of 3 groups of bursts.\\
The first group is composed of 19 \emph{HETE}-bursts with redshift.
This group has a cut in its redshift distribution at z = 3.3
\textit{(figure \ref{tab2}, dotted line)}. Nevertheless, this cut
seems to disappear and for the second group composed of 53
\emph{HETE}-GRBs with redshift or pseudo-redshift \textit{(solid
line in figure \ref{tab2})}, the cumulative distribution of this
group is fully compatible with the group of 22 \emph{SWIFT}-GRBs
with measured redshift \textit{(figure \ref{tab2}, dashed line)}. This result tends to show that the redshift-distribution of
\emph{HETE}-GRBs is biased at high-redshift.\\
Finally, we note that the sample studied contains few high-redshift
GRBs : 4 GRBs only have a redshift higher than z = 4
(\emph{GRB010612, 030913, 031026} and \emph{051008}).

\subsection{Comments on the \emph{npz} and possible applications}
Considering the most intense part of the GRBs seems a good way to
improve the pseudo-redshift based on the prompt emission. Indeed, if
we consider some examples such as \emph{GRB010612} and
\emph{GRB031026}, they were previously found at $\widehat{z} = 9.5$
and $14$ \cite{Atteia04}, compared to the \emph{npz} which now gives
respectively $5.3$ and $6.7$, values probably closer to the real redshift.\\
Moreover, it has solved the problem of \textit{multi-peaks} GRBs in
which the background was taken into account in the determination of
$t_{90}$, which had for consequence a biased value of $X_{0}$. For
example, \emph{GRB020305} had an estimation of $5.88$
\cite{Atteia04}, and is now found at $1.98\pm1.45$, in agreement
with the spectroscopic constraints ($z \leq 2.8$, \cite{Goros}).\\
Finally, if we consider recent determinations of spectroscopic
redshift for old bursts, we can notice that \emph{GRB030528}
($z=0.782$, \cite{Rau}) has a close $npz$ of $0.64$. For
\emph{GRB020819}, we find a value of $npz = 1.21$, not close to the
real redshift ($z = 0.41$, \cite{Jakob}), but the error ($\pm~0.9$)
which is also large makes the estimation compatible with the true
redshift.\\
The development of the pseudo-redshift finds several possible applications.\\
As the pseudo-redshift are rapidly computed, they can tell us very
quickly whether the burst is at low or high-redshift, which permits to choose the appropriate way of observation.\\
\\
Other applications of pseudo-redshift could be the verification of
the validity of the $E_{p}-E_{iso}$ relation found by Amati
\cite{Amati02} for a large sample of GRBs \cite{Pizz}.\\
Finally, having a large sample of GRBs with redshift (or estimation)
should let us study some of their cosmological aspects such as their
luminosity function or the evolution of their rate with the
redshift.

\end{document}